\documentclass{article}

\usepackage{siunitx}
\usepackage{multirow}
\usepackage{authblk}
\usepackage{graphicx}
\usepackage{geometry}
\usepackage{url}
\usepackage{enumitem}

\begin{document}
\title{Sampling requirements in near-field ptychography}

\author[1*]{Luca Fardin}
\author[2,3]{Yelyzaveta Pulnova}
\author[2,4]{Tom\'a\v s Parkman}
\author[2,3]{Iuliia Baranov\'a}
\author[5]{Sylvain Fourmaux}
\author[6]{Chris Armstrong}
\author[7,8]{Michela Fratini}
\author[2]{Uddhab Chaulagain}
\author[2,4]{Jaroslav Nejdl}
\author[2]{Borislav Angelov}
\author[9]{Darren J. Batey}
\author[1]{Alessandro Olivo}
\author[1]{Silvia Cipiccia}

\affil[1]{\raggedright\small\itshape{Department of Medical Physics and Biomedical Engineering, University College London, Malet Place Engineering, Gower St, London, WC1E 6BT, UK}}
\affil[2]{\raggedright\small\itshape{The Extreme Light Infrastructure ERIC, ELI Beamlines Facility, Za Radnicí 835, 252 41 Dolní Břežany, Czech Republic}}
\affil[3]{\raggedright\small\itshape{Charles University, Faculty of Mathematics and Physics, Ke Karlovu 3, 121 16 Prague, Czech Republic}}
\affil[4]{\raggedright\small\itshape{Czech Technical University in Prague, Faculty of Biomedical Engineering, nam. Sitna 3105, 27201 Kladno, Czech Republic}}
\affil[5]{\raggedright\small\itshape{Institut National de la Recherche Scientifique—Énergie, Matériaux et Télécommunications, Université du Québec, 1650 Lionel Boulet, Varennes, J3X 1P7, QC, Canada}}
\affil[6]{\raggedright\small\itshape{Department of Plasma Physics, Central Laser Facility, RAL. OX11 0QZ, United Kingdom}}
\affil[7]{\raggedright\small\itshape{Institute of Nanotechnology, CNR-Nanotec, Piazzale Aldo Moro 5, 00185, Rome, Italy}}				
\affil[8]{\raggedright\small\itshape{Santa Lucia Foundation, via Ardeatina 356, 00179, Rome, Italy}}
\affil[9]{\raggedright\small\itshape{Diamond Light Source, Harwell Science and Innovation Campus, Fermi Avenue, Didcot, OX11 0DE, UK}}
\affil[*]{\raggedright\small\itshape{l.fardin@ucl.ac.uk}}

\date{}
\maketitle

\begin{abstract}

Ptychography is a robust lensless form of microscopy routinely used for applications spanning life and physical sciences. The most common ptychography setup consists in using a detector to record diffraction patterns in the far-field. A near-field version has been  more recently introduced, and its potential is yet to be fully exploited. In this work, the sampling requirements for near-field ptychography are analysed. Starting from the characterisation available in literature, the formalism of the fractional Fourier transform is used to generalise analytically the sampling conditions. The results harmonise the far- and near-field regimes and widen the applications of the technique with respect to the current knowledge. This study is supported by simulations and provides clear guidelines on how to optimise the setup and acquisition strategies for near-field ptychography experiments. The results are key to drive the translation of the technique towards low brilliance sources. 

\end{abstract}

\section{Introduction}

 Ptychography is a coherent diffraction imaging (CDI) technique that simultaneously provides information on the modulus and phase of the sample (object) and illumination (probe) \cite{rodenburg_07}. It has been applied successfully to the electron\cite{Nellist:sp0116}, visible\cite{Wang_2023} and x-ray regime\cite{PhysRevLett.98.034801}. In this work we focus on the latter. Ptychography does not use any optics between the sample and the detector to form the image and is therefore free from any lens aberrations, giving access to diffraction limited resolutions.  A conventional ptychography acquisition consists in recording diffraction patterns in the far-field while scanning the sample across a coherent illumination, at overlapping steps. In standard CDI, the maximum size of the illumination is determined and limited by the far-field condition \cite{paganin_book_2006}. The scanning process provides diversity in the diffraction patterns, as at each step new regions of the sample are illuminated. If sufficient overlap is maintained during the scanning \cite{edo_sampling_2013,silva_sampling_2015}, the information-rich dataset enables an iterative retrieval algorithm \cite{Marchesini_2016,Enders_2016,Maiden_2009,Guizar_2008,Wakonig_2020} to decouple the transmittance function of the sample from the illumination, and to solve for the phase problem. The technique is fully quantitative, as the retrieved sample phase  is directly proportional to its electron density. 
 The drawback of the technique is the need for a larger number of diffraction patterns, and therefore lengthy scans, especially for extended objects. 

To reduce the acquisition time for large samples, Stockmar et al. extended ptychography to the near-field regime \cite{stockmar_near-field_2013}. In near-field ptychography (NFP), the size of the illumination is not limited by the far-field condition, therefore NFP can be used to image larger objects with fewer diffraction patterns  \cite{stockmar_near-field_2013,clare_characterization_2015} compared to far-field ptychography (FFP).  However, this comes at the cost of a reduced spatial resolution, which, in the near-field, is determined by the (demagnified) pixel size of the imaging system. In NFP the diversity between diffraction patterns is obtained by structuring the illumination with speckles produced with a beam modulator.

In the X-ray regime, NFP has been successfully performed at synchrotron facilities, both in 2-D and 3-D \cite{monaco_comparison_2022,stockmar_ptychotomo_2015}. Compared to other full-field coherent imaging techniques, such as in-line holography\cite{Guizar-Sicairos_2007}, while NFP requires more computational power, it is robust against highly absorbing and optically thick samples \cite{stockmar_thick_2015}, and has been proven to produce comparable electron density maps \cite{monaco_comparison_2022}. 
Moreover, unlike in-line holography, the reconstruction algorithms do not require prior knowledge of the composition of the sample, making NFP more versatile for quantitative analysis \cite{monaco_comparison_2022}.

As NFP has been proposed relatively recently (2013, \cite{stockmar_near-field_2013}), there is little available in the literature on its characterisation, beyond the guidelines provided by Clare et al. \cite{clare_characterization_2015}. Clare and collaborators carried out an extensive investigation on the influence on NFP of different parameters including, but not limited to, the beam modulator properties, the scanning step size and the propagation distances between modulator, sample, and detector. The interpretation of the results was given in terms of local interference between the structured illumination and the diffraction pattern of the sample, occurring for each scanning position within the first Fresnel radius \cite{clare_characterization_2015}, where the $k^{th}$ Fresnel radius at a distance $d$ is defined as $R_k=\sqrt{k\lambda d}$, with $\lambda$ the X-ray wavelength. 
To guarantee sufficient diversity between scanning points for a good convergence of the reconstruction algorithm, it was concluded that the step size had to be greater than the speckle size, as well as the largest feature sizes in the sample. Moreover,  the propagation distance had to be large enough to provide good speckle visibility, and to ensure the first Fresnel radius was larger than the speckles.
Based on the symmetry between object and illumination reported by Clare \cite{clare_characterization_2015}, this would imply that also the feature sizes of the object should be of the order of the first Fresnel radius, a condition satisfied in most of the previous works on X-ray NFP \cite{stockmar_near-field_2013,stockmar_ptychotomo_2015,clare_characterization_2015,monaco_comparison_2022}. It was also reported that large homogeneous objects produce weak interference fringes at their center and no coupling with the speckles of the illumination \cite{stockmar_near-field_2013}. Low-frequency artefacts in the retrieved sample phase were therefore expected  \cite{stockmar_near-field_2013}, and experimentally observed \cite{clare_characterization_2015} at the center of larger structures. This would imply that NFP is not suitable to image samples containing relatively large homogeneous features. 

In this paper we investigate the sampling conditions for NFP to understand if and how these limitations can be overcome. We generalise the previous studies with an analytical approach supported by simulations. Based on the findings, we provide guidelines to deal with different samples and experimental geometry and in doing so, increase the applicability of NFP both within synchrotron facilities, and toward laboratory applications.

\section{Sampling conditions in NFP}
In order to derive the sampling conditions in the case of near-field propagation, the far-field case is briefly reviewed.
Considering the one dimensional case for simplicity, and assuming the sample is thin enough to neglect the diffraction within the scattering volume (projection approximation), a monochromatic wavefront $\Psi_0(x^{\prime})$ at the exit surface ($x^{\prime}$ plane) of the object can be expressed as:
\begin{equation}
\Psi_0(x^{\prime})=P(x^{\prime})O(x^{\prime})
\end{equation}
where $O(x^{\prime})$ is the complex object transfer function and $P(x^{\prime})$ is the complex illumination (probe). The free-space propagator to transport the exit wave from the sample to a detector in the far-field, can be approximated by a Fourier transform. Therefore, using the scaled Fourier transform formalism \cite{paganin_book_2006}, the transported wave at the detector plane $x$, located at a distance $z=d$ from the sample is given by:
\begin{equation}\label{eq:FFP_propagation}
    \Psi_D(x) = \mathcal{F}\left[P(x^{\prime})O(x^{\prime})\right]\left(\frac{x}{\lambda d}\right)
\end{equation}
where $\mathcal{F}$ denotes the Fourier transform and $\lambda d$ is the scaling factor. 

This formalism was used to derive the real space sampling condition in ptychography by starting from a discrete windowed Fourier transform representation\cite{silva_sampling_2015}, and from the Nyquist-Shannon sampling theorem \cite{batey_th_2014}. Let us first introduce a fundamental quantity in CDI, i.e. the maximum reconstruction window $W$, which, for a detector pixel size $\Delta x$, is given by:
\begin{equation}\label{eq:window_CDI}
    W = \frac{\lambda d}{2 \Delta x} 
\end{equation}
This physically represents the maximum spatial extent of the illumination that can be used in CDI at the object plane, i.e. the probe width. In FFP, the real space sampling condition can be expressed in terms of W as a maximum scanning step size $\mathrm{\Delta_{step}}$ \cite{batey_th_2014}:
\begin{equation}\label{eq:sampling_ffp}
    \mathrm{\Delta_{step}} < W
\end{equation}
which is valid when the number of scanning steps N $\gg 1$.
As described by Batey \cite{batey_th_2014}, each scanning step in ptychography extends $W$ of $\mathrm{\Delta_{step}}$, thus creating an extended (synthetic) illumination $W_S = W + (N-1) \mathrm{\Delta_{step}}$ after N steps, that corresponds to the field of view of the scan.

In the near-field regime, the far-field approximation of eq. \ref{eq:FFP_propagation} cannot be applied, and the propagator needs to be differently formulated. It has been shown that this can be described via a scaled fractional Fourier transform (FrFT) \cite{Pellat-Finet,Ozaktas:95}. FrFT is a generalisation of the Fourier transform, defined as \cite{Almeida:94}:
\begin{equation}\label{Eq:FrFT}
    \mathcal{F}^{\alpha}[f](u) = \sqrt{1-i\cot{\alpha}} \int_{-\infty}^{\infty} \exp\left[ i\pi(u^2 \cot{\alpha} -2ux^{\prime}\csc{\alpha}+x^{\prime 2}\cot{\alpha}) \right] f(x^{\prime}) \mathrm{d}x^{\prime}
\end{equation}
The parameter $\alpha$ can be interpreted as the angle between the real and the frequency space: when $\alpha=0$ the FrFT coincides with the identity function, for $\alpha=\pi/2$ the FrFT is equal to the Fourier transform, and for $\alpha=-\pi/2$ the FrFT is equal to the inverse Fourier transform. 
Using the FrFT formalism the near-field propagation can be expressed as \cite{Ozaktas:95,Ozaktas:11}:
\begin{equation}\label{eq:nfp_frft}
    \Psi_D(\tilde{x}) = e^{\frac{i2\pi d}{\lambda}}e^{-\frac{i\pi \alpha}{4}} \sqrt{\frac{1}{M}} \exp{\left( \frac{i\pi x^2}{\lambda \rho}\right)} \mathcal{F}^{\alpha}[\Psi_0(\tilde{x}^{\prime})]\left(\frac{\tilde{x}}{M}\right)
\end{equation}
\begin{equation}\label{eq:mscale}
    M = \sqrt{1+\frac{\lambda^2 d^2}{s^4}}
\end{equation}
\begin{equation}\label{eq:lambda_prop}
    \alpha = \arctan{\frac{\lambda d}{s^2}}
\end{equation}
\begin{equation}
    \rho = d \csc{\alpha}
\end{equation}
where $\tilde{x}=x/s$ and $\tilde{x}^{\prime}=x^{\prime}/s$ are adimensional variables, and $s$ an arbitrary length constant. The quadratic phase term outside the scaled FrFT in eq. \ref{eq:nfp_frft} can be neglected when measuring the intensity:
\begin{equation}\label{eq:FrFT_intensity}
   \left | \Psi_D(\tilde{x})\right | ^2 = \frac{1}{M}  \left | \mathcal{F}^{\alpha}[\Psi_0(\tilde{x}^{\prime})]\left(\frac{\tilde{x}}{M}\right) \right | ^2
\end{equation}

The Nyquist-Shannon theorem for the Fourier transform can be extended to a function $f(\tilde{x}^{\prime})$ band-limited in $[-\Omega/2,\Omega/2]$ in the FrFT domain \cite{Xia:96}, and it is expressed by:
\begin{equation}\label{eq:shannon}
   \Delta \tilde{x}^{\prime} \leq \frac{\sin{\alpha}}{\Omega}
\end{equation}
$\Delta \tilde{x}^{\prime}$ is the sampling interval of $f(\tilde{x}^{\prime})$ that, in NFP, coincides with the  detector pixel $\Delta \tilde{x}$. By using eq. \ref{eq:mscale} and \ref{eq:lambda_prop} and by switching back to dimensional variables, eq. \ref{eq:shannon} can be rearranged as:
\begin{equation}\label{eq:window_width}
    s \Omega M \leq \frac{\lambda d}{\Delta x}
\end{equation}
The quantity $s\Omega M$ has the dimension of a length and it represents the maximum extent in the detector plane beyond which the diffraction interference fringes cannot be resolved by sampling $\left | \Psi_D\right | ^2$ with $\Delta x$. Indeed, the maximum frequency interval that can be sampled with pixel size $\Delta x$ is $1/\ \Delta x$. This corresponds to an outgoing divergence of $\theta = \lambda/\ \Delta x$ and to a spread of the diffraction at distance $d$ of $\lambda d /\ \Delta x$ \cite{Ozaktas:11} as in eq. \ref{eq:window_width}. 

\begin{figure}[htbp]
\centering\includegraphics[scale=0.25]{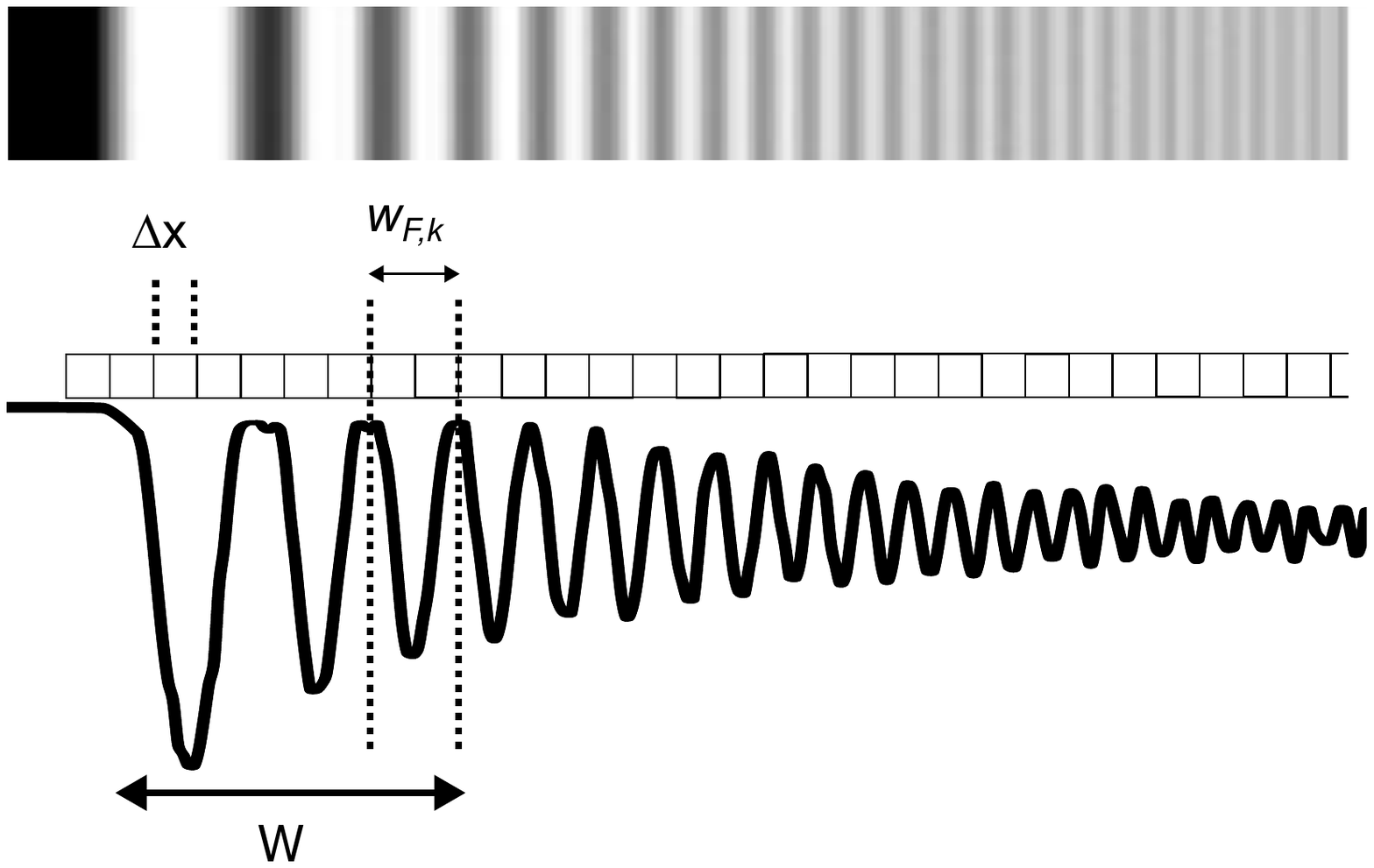}
\caption{Schematic representation of a diffraction from an edge sampled with a detector pixel size $\Delta x$. The largest extent that can be resolved is $W$.}
\label{fig:diffraction}
\end{figure}

Another way to derive the same condition via a more intuitive approach is by considering the diffraction from an edge after a propagation distance $d$. Let us approximate the separation $w_F$ between two adjacent minima/maxima of the diffraction pattern, by using of the expression for the $k^{th}$ Fresnel radius ($R_k$): $w_{F,k} = R_{k}-R_{k-2}$ , $R_k = \sqrt{k \lambda d}$. The smallest interference fringe that can be detected with a detector pixel $\Delta x$ corresponds to the largest $k$ such that $w_{F,k} > 2\Delta x$, that leads to (k>>1)  $\sqrt{k} < \sqrt{\lambda d} /\ 2 \Delta x$ (see Figure \ref{fig:diffraction}). Since in NFP a complex function is estimated from real valued diffraction patterns, 
an additional factor 2 is expected to appear at the denominator in the right-hand-side of eq. \ref{eq:window_width}.
This corresponds to a diffraction pattern size of $2 R_k \approx \lambda d /\ \Delta x$, as found in eq. \ref{eq:window_width}. 

Summarising, the following NFP sampling conditions are introduced:
\begin{enumerate}[label=c.\arabic*,ref=c.\arabic*]
    \item \label{1}$\mathrm{W} = \lambda d /\ 2\Delta x$ is the window of interest in NFP. This corresponds to the sampling condition for a compact support in FFP (eq. \ref{eq:window_CDI}). A maximum $\mathrm{\Delta_{step}} < \mathrm{W}$ can be used to ensure correlation between steps.
    \item \label{2} The process of scanning extends W of the scanning step $\mathrm{\Delta_{step}}$, thus creating a local synthetic window $W_S$: $W_S=W+(N-1)\Delta_{step}$, as for FFP. If the object $O(x)$ has features at a distance $\mathrm{w_o}$ and is uniform in between them, in order to avoid artefacts in the reconstruction (as in \cite{clare_characterization_2015}), the number of scanning steps $N$ has to satisfy the condition: $W_S \geq \mathrm{w_o}$.
    
\end{enumerate}

In the following section, we examine and verify the conditions \ref{1} and \ref{2} via wave-optics simulations.

\subsection{Simulations}

\begin{figure}[htbp]
\centering\includegraphics[scale=0.5]{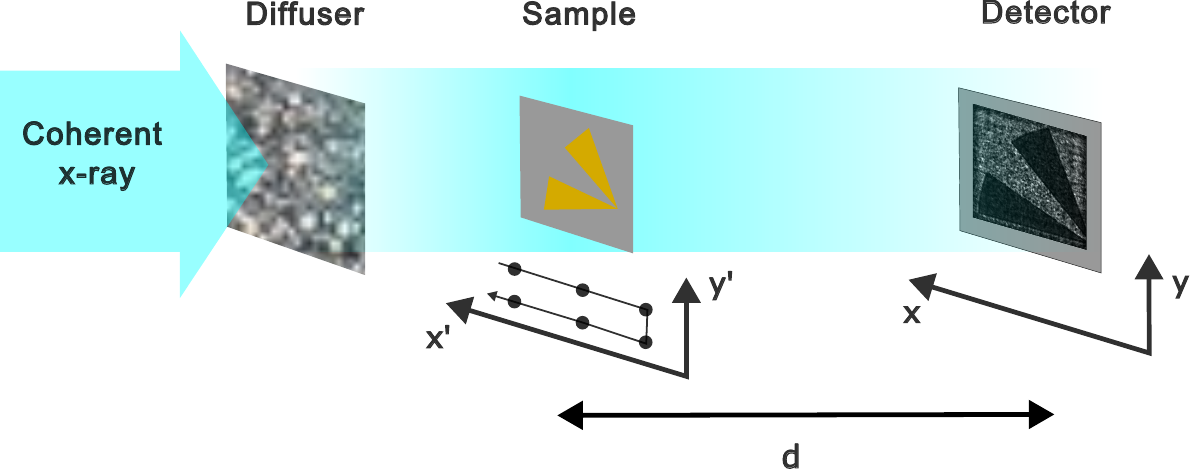}
\caption{Graphical representation of the simulation setup in parallel geometry. The diffuser-sample distance is 2 m. The sample-detector distance is varied as reported in Table \ref{table:Figure_step_summary}.}
\label{fig:exp_set}
\end{figure}

For simplicity, all the simulations in this section are in parallel geometry: the Fresnel scaling theorem ensures no loss of generality \cite{paganin_book_2006}. All the simulations of the NFP diffraction patterns are based on the python library Diffractio\cite{diffractio}. The simulated setup is shown in figure \ref{fig:exp_set}: a fully coherent illumination is structured with a random pattern modulator, propagated to the sample object and from the sample to a detector in the near-field, by using the Diffractio's chirped-z transform propagator \cite{bluestein_czt,hu_efficient_2020}. The X-ray energy is \qty{8}{\kilo\electronvolt} and the simulated detector pixel size is \qty{0.42}{\micro\meter}, unless otherwise stated. This value was chosen to be at least one order of magnitude smaller than the first Fresnel radius. The average size of the speckles, measured as the full-width at half-maximum of the autocorrelation function, is \qty{2}{\micro\meter}. The sample is a Siemens star test pattern simulated with 50\% transmission (modulus) and \qty{-1}{\radian} (phase). The spokes are generated by dividing a circle into equal angular intervals, whose amplitude controls the size $w_o$ of the object features (see Supplementary material, figure S1). The object is translated on a 2-D regular grid of $N \times N$ positions: although this pattern may give rise to well-known periodic reconstruction artefacts \cite{Thibault_2009}, it defines unambiguously the scanning step size and range.
The ptychography reconstructions have been performed using the implementation of the ePIE algorithm\cite{maiden_ePIE_2009} in PtyREX code\cite{batey_th_2014}.
The quality of the reconstructions was evaluated by using the root mean square error (RMSE) between the object's phase retrieved with PtyREX and the ground truth simulated with Diffractio.

\section{Results}
\begin{figure}[h]
\centering\includegraphics[scale=0.26]{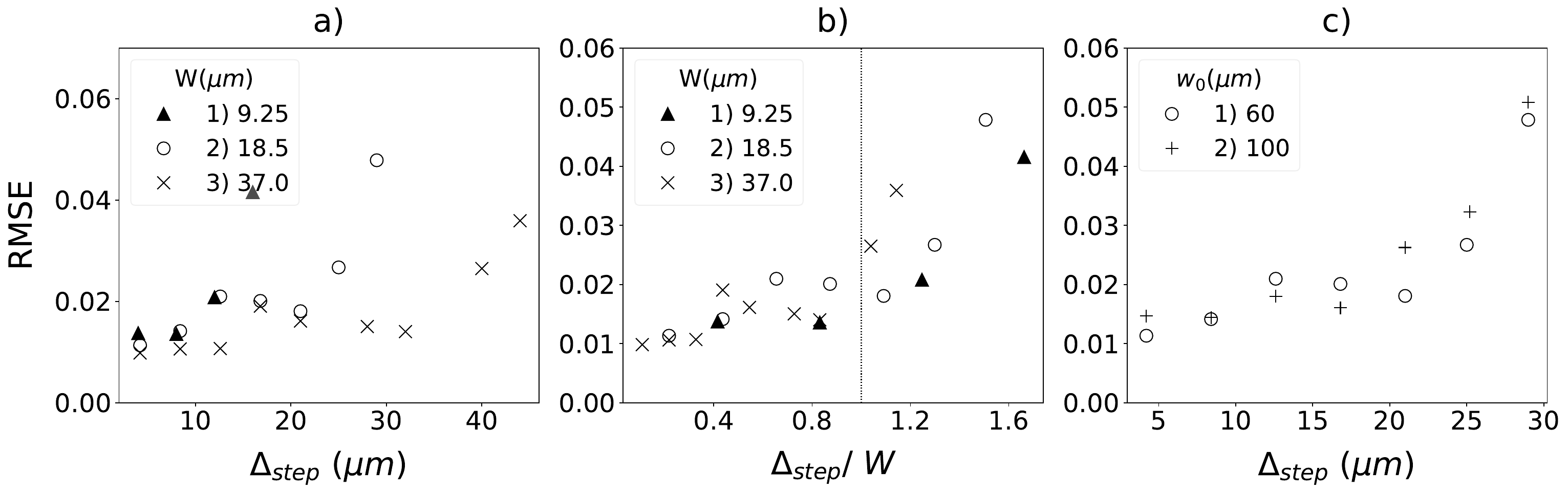}
\caption{Effect of the scanning step over the reconstructed image quality for $O_1$. RMSE for different $W$ as a function of: (a) the scanning step in real unit and (b) normalised by $W$. Vertical dotted line: $\Delta_{step}/\ W=1$ . In (c), two different outermost spokes widths $O_1$ and $O_2$ are compared while keeping $W$ constant.}
\label{fig.figure_step}
\end{figure}

In the first set of simulations we investigate the effect of the step size on the quality of the retrieved images. The analysis is performed while always satisfying the sampling condition \ref{2}, i.e. by keeping the synthetic window $W_S$ at least as large as the width of the object features $w_o$. Figure \ref{fig.figure_step}-a shows the RMSE as a function of the step size $\Delta_{step}$ in \unit{\micro\meter}, for different widths of the diffraction pattern $W$, while keeping $w_o$ constant. $W$ is varied by changing the detector pixel size and the sample-detector distance (see the simulation parameters for figure \ref{fig.figure_step} summarised in table \ref{table:Figure_step_summary}). We can see that the RMSE diverges for smaller step sizes as $W$ becomes smaller. This is compatible with the sampling condition \ref{1}. To further clarify this, figure \ref{fig.figure_step}-b presents the same results by plotting RMSE versus the step size normalised by $W$. In this case, the curves tend to overlap and the RMSE diverges when $\Delta_{step} /\ W>1$, confirming condition \ref{1}. Finally, figure \ref{fig.figure_step}-c shows the RMSE versus the step size for two different $w_o$ (\qty{60}{\micro\meter} and \qty{100}{\micro\meter} respectively for $O_1$ and $O_2$). The two curves overlap, confirming that the maximum step size does not dependend on the object features. 

\begin{table}[ht]
\fontsize{8pt}{8pt}{
        \centering
        \tabcolsep=0.18cm
\small
\caption{Summary of the simulations parameters\\ for figure \ref{fig.figure_step} and figure \ref{fig.figure_range}.}
        \begin{tabular}{c|c c c c c   }
 \hline
&Dataset&$w_o$[\unit{\micro\meter}] &$\Delta$ x[\unit{\micro\meter}] & $d$ [\unit{\centi\meter}] & $W$ [\unit{\micro\meter}]\\
\hline
\multirow{5}{1em}{\rotatebox{90}{Figure \ref{fig.figure_step}}}&a/b-1&$60$&$0.83$&$10$& $9.24$\\
&a/b-2&$60$&$0.42$&$10$& $18.5$\\
&a/b-3&$60$&$0.42$&$20$& $37.0$\\
\cline{2-6}
&c-1&$60$&$0.42$&$10$& $18.5$\\
&c-2&$100$& $0.42$&$10$& $18.5$\\
\hline
 \hline
\multirow{3}{1em}{\rotatebox{90}{Figure \ref{fig.figure_range}}} &a/b-1   & $60 $   &0.42 & $10 $& $18.5$\\
 &a/b-2   & $60 $   &0.42 & $20 $& $37.0$\\
 &a/b-3   & $100 $  &0.42  & $10 $& $18.5$\\

\end{tabular}

\label{table:Figure_step_summary}
}
\end{table}
   
\begin{figure}[htbp]
\centering\includegraphics[scale = 0.26]{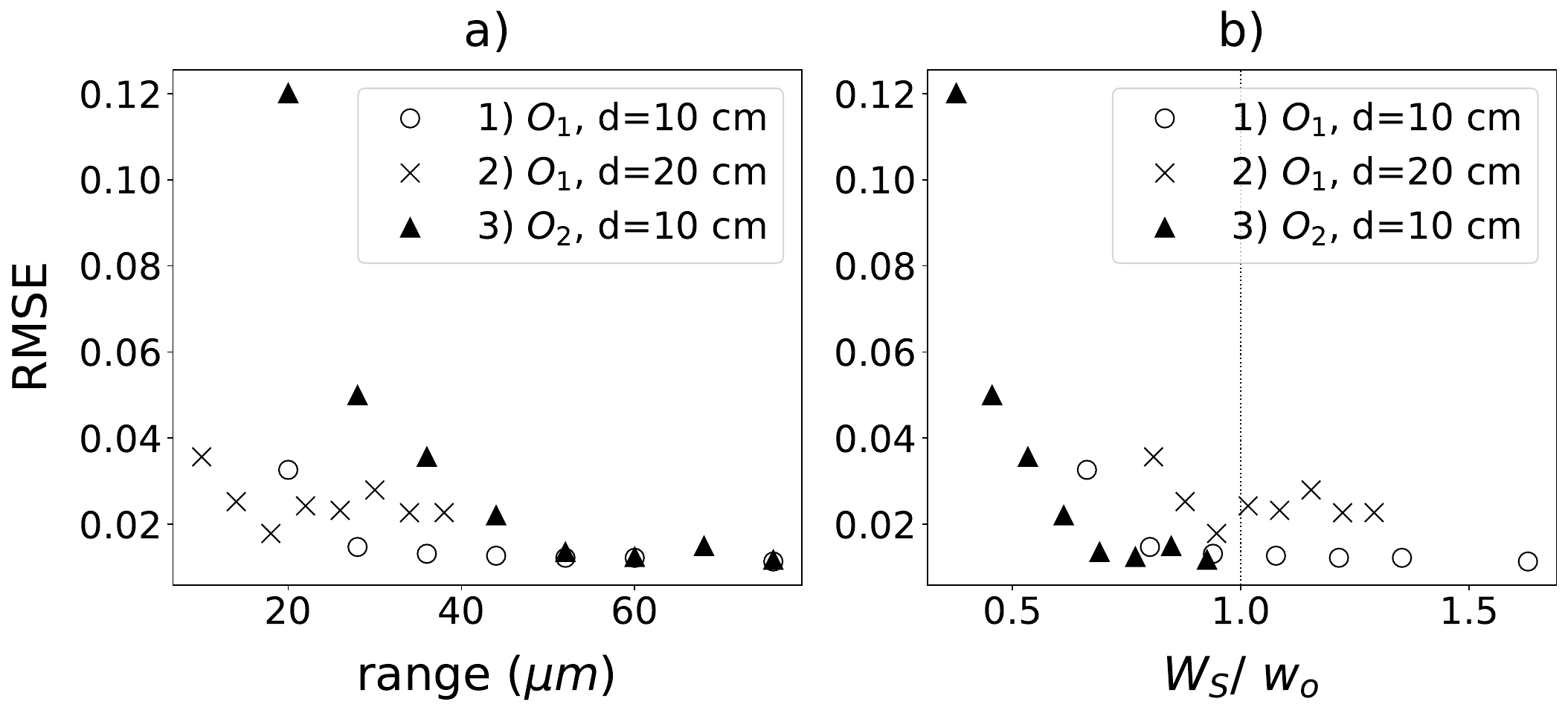}
\caption{Effect of the scanning range over the reconstructed image quality. RMSE as a function of: a) the scanning range in real units and b) $W_S/\ w_o$. The vertical dotted line indicates a ratio of 1. The step size is \qty{4}{\micro\meter}, except for the propagation distance $d=$\qty{20}{\centi\meter}, where it was reduced to \qty{2}{\micro\meter}}.
\label{fig.figure_range}
\end{figure}

The validity of condition \ref{2} is analysed in the second set of simulations. The quality of the reconstructions is evaluated via the RMSE as a function of the total scanning range, defined as  $(\mathrm{N}-1) \Delta_{step}$, while varying $w_o$ and the diffraction window $W$. The step size is kept constant to \qty{4}{\micro\meter} for $d=$ \qty{10}{\centi\meter}, and is reduced to \qty{2}{\micro\meter} for $d=$ \qty{20}{\centi\meter}, in order to keep $\mathrm{N} > 6$ for every configuration under study. In both cases the step was chosen to be well within the range of validity of the sampling condition \ref{1} (figure \ref{fig.figure_step}). Figure \ref{fig.figure_range}-a shows the results: by increasing $w_o$ or reducing $d$ (i.e. $W$), a larger scanning range is required to improve the reconstruction's RMSE. Figure \ref{fig.figure_range}-b shows the same results when the scan range is expressed in terms of the synthetic reconstruction window $W_S$ normalised against $w_o$. The reconstruction always converges when $W_S/\ w_o>1$ and becomes less accurate as $W_S/\ w_o$ decreases below 1. 

\begin{figure}[htbp]
\centering\includegraphics[scale = 0.3]{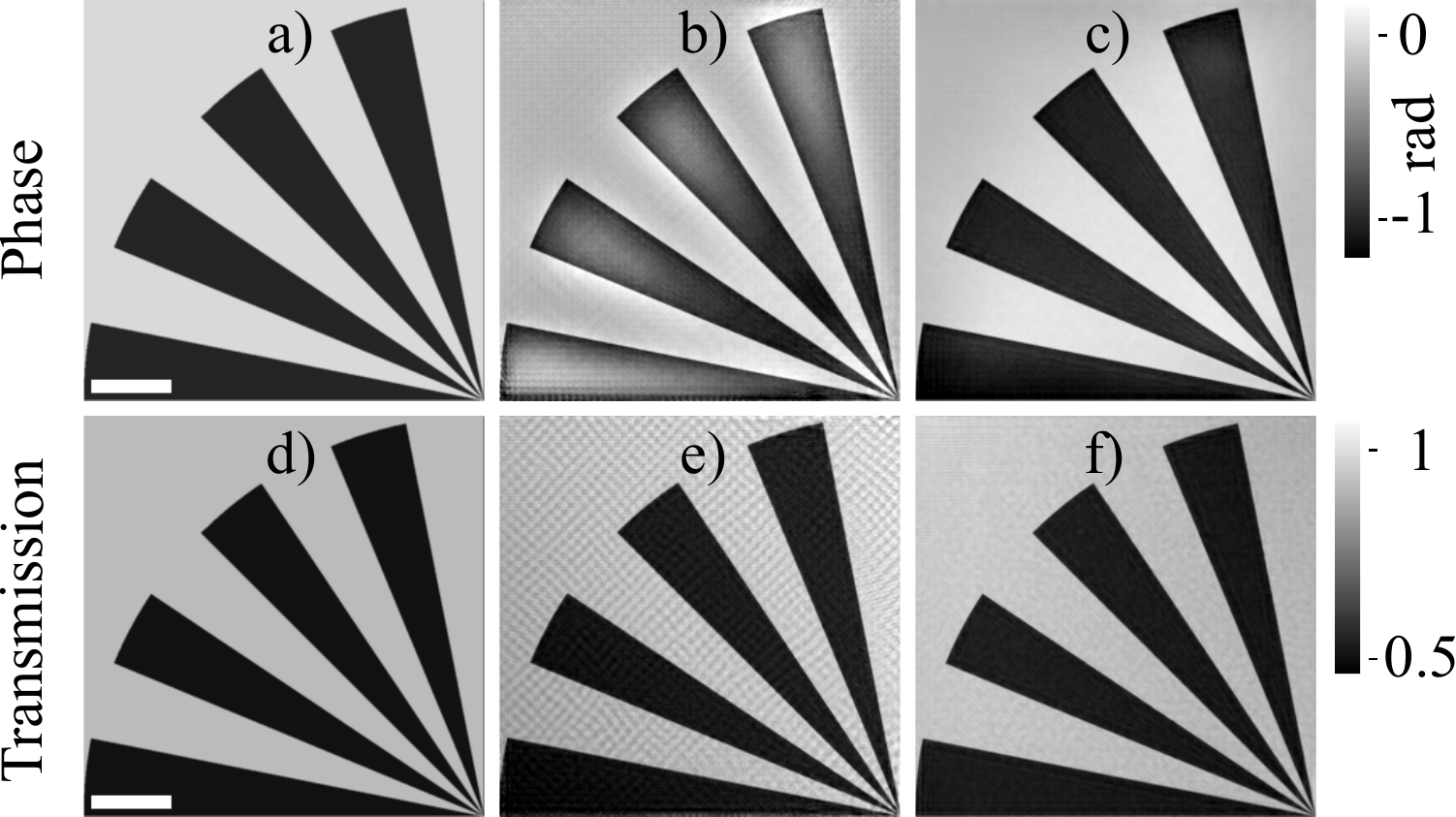}
\caption{Object phase (a-c) and modulus (d-f) reconstructed from simulations. (a,d) ground truth object ($O_1$, $w_0 = $ \qty{60}{\micro\meter}). (b,e) reconstruction obtained from $d=$\qty{10}{\centi\meter}, $\Delta_{step}=$ \qty{4}{\micro\meter} and $W_S$ = \qty{30.5}{\micro\meter}. (c,f) same as (b,e) with $W_S$ = \qty{78.5}{\micro\meter}. (b,e) correspond to an insufficient scanning range and show artefacts in the reconstructed phase. The scale bars are equal to \qty{50}{\micro\meter}.}
\label{fig.figure_artefact_range}
\end{figure}

This is shown also in Figure \ref{fig.figure_artefact_range} where (a) and (d) are the phase and modulus of the ground truth object $O_1$ ($w_o=$ \qty{60}{\micro\meter}) simulated with Diffractio, (b) and (e) are the reconstructions obtained with a $W_S$ of \qty{30.5}{\micro\meter} , and (c) and (f) with a $W_S$ of \qty{78.5}{\micro\meter}.
To notice that the signature of an NFP scan not satisfying the sampling condition \ref{1} or \ref{2} is a low-frequency artefact in the retrieved object's phase, which appears hollow in between features as in (b), where $W_S/\ w_o < 1$ and the phase tends towards the background value. This is particularly visible moving away from the center of the star, where the features' distance increases. The modulus is less affected by the violation of the sampling conditions, even if background artefacts appear in (e), associated to a non-converging phase reconstruction in (b). Equivalent low-frequency artefacts appear when the step size is too large and the first sampling condition is not respected: this is shown in the Supplementary materials (figure S2). Similar artefacts in the phase reconstruction were observed experimentally when the NFP sampling conditions were not satisfied: the measurement and associated results are shown in the Supplementary materials (figures S3 and S4). 

\section{Discussion and Conclusions}
We have investigated the sampling conditions in NFP, starting from the pioneering works of Stockmar\cite{stockmar_near-field_2013} and Clare\cite{clare_characterization_2015}. We have defined a window of interest $W$, in terms of the Fresnel radius squared and the pixel size. This is summarised by equation \ref{eq:window_width} and sampling condition \ref{1}, which is the equivalent of the sampling condition for a compact support in the far-field regime (eq. \ref{eq:window_CDI}). 
We have found that, as for the far-field case, the  maximum step size has to be smaller than the window $W$ to achieve convergence. It should be noticed that in the NFP case the window $W$ 
 is defined at the detector plane, while in the FFP it is defined in the object plane. A comparison between the two cases can be performed based on the properties of near-field diffraction: going from the sample to the detector plane, the pixel size and the translation distances are not rescaled. Therefore, $W$ imposes in NFP an equivalent condition as in FFP on the theoretical maximum step size in the object plane \cite{edo_sampling_2013,silva_sampling_2015}. Experimentally, in FFP  a more stringent condition was found, with an overlap in the illumination of at least $60\%$ in between steps. The simulations in this work were performed in ideal conditions: effects of noise, instability of the setup and partial coherence (among others) were not investigated. An experimental validation of the sampling conditions will be the subject of a future study.

Following the considerations by Batey \cite{batey_th_2014}, we have introduced a synthetic window $W_S$ which is equivalent to the synthetic illumination in FFP and we have shown that, if it is at least as large as the maximum width of the objects' features $w_o$, an accurate reconstruction is obtained (in terms of RMSE). As $W_S$ decreases below $w_o$, artefacts appear in the reconstruction, as observed in figure \ref{fig.figure_artefact_range}-b, and in previous works \cite{stockmar_near-field_2013,clare_characterization_2015}.

In published works on NFP, the feature sizes of the objects were always on the order of a few microns, (e.g. the Siemens Star spokes in \cite{stockmar_near-field_2013,clare_characterization_2015} and the mean pore diameters in \cite{stockmar_ptychotomo_2015,monaco_comparison_2022}), compared to a window $W$ always greater than \qty{10}{\micro\meter}. Therefore, in all the publications analysed, condition \ref{2} was satisfied. Only in one previous publication, the parameters were close to violating condition \ref{2}. In reference \cite{stockmar_thick_2015} a \qty{46}{\micro\meter} uranium sphere was imaged  with $W=$ \qty{23}{\micro\meter} and $W_S \approx$ \qty{45}{\micro\meter}: due to the presence of strong phase gradients when moving towards the edges, this sample can be considered homogeneous only around the center, where indeed the error in the phase was greater.

An analysis of the maximum step size was performed numerically by Clare et al. \cite{clare_characterization_2015}. The authors used $4 \times 4$ scanning positions randomly distributed over different total scanning ranges. The diffraction window in their work was $W=$ \qty{23}{\micro\meter}. They found that the RMSE started diverging for scan ranges larger than \qty{80}{\micro\meter}, explained in the paper as being due to the sample going out of the field of view (FoV). However, in all the simulations presented in our work, the sample is almost as large as the FoV and partially goes out of it, upon translation. For example, in figure \ref{fig.figure_step}, for a propagation distance of \qty{20}{\centi\meter}, convergence was obtained for a step size up to \qty{30}{\micro\meter}: in this configuration, at the edges of the scanning range (N=6), 50\% of the sample is outside the field of view. This suggests that NFP can reach convergence even when the sample leaves the FoV. The increase in RMSE observed by Clare \cite{clare_characterization_2015} for translation ranges larger than \qty{80}{\micro\meter} can be explained in terms of violation of sampling condition \ref{1}. Indeed for ranges $>$ \qty{80}{\micro\meter} the average step size is $\Delta_{step}>$ \qty{25}{\micro\meter}, that is $\Delta_{step}> W$. 

In the other aforementioned experimental works on X-ray NFP \cite{stockmar_near-field_2013,stockmar_ptychotomo_2015,stockmar_thick_2015,monaco_comparison_2022}, the scanning points are instead distributed over a pseudo-random grid, which differs from a regular lattice. Due to this configuration, a direct comparison with our work is not possible. However, the total scanning range was always found to be close to the experimental $W$, thus satisfying condition \ref{1}.

In conclusion, by analysing diffraction in the near-field regime with the help of the FrFT formalism, we have identified an analytical expression for the maximum step size in NFP and the minimum scanning range based on the largest sample features. These conditions find correspondence in far-field ptychography, harmonising the two regimes. Our investigation has been focused on X-ray imaging. However, we expect the findings to be applicable to other radiation regimes. The results provide general guidelines to design optimised experimental setups and acquisition strategies for NFP. This is expected to facilitate any future transition of the technique from bright sources (e.g. synchrotrons) to laboratory scale facilities.

\section*{Acknowledgments}
This work is supported by EPSRC New Investigator Award EP/X020657/1 and and Royal Society RGS/R1/231027. A.O. is supported by the Royal Academy of Engineering under the Chairs in Emerging Technologies scheme (grant CiET1819/2/78). We thank the ELI beamlines for the experimental beamtime under the proposal ELIUMP-91. We would like to thank Julio Cesar da Silva and Andrea Mazzolani for the fruitful discussions. 

\section*{Supplemental document}
See Supplement 1 for supporting content.

\bibliography{nfp}
\bibliographystyle{ieeetr}

\end{document}


\maketitle

\section{Simulated sample}

\begin{figure}[h]
\centering\includegraphics[scale=0.75]{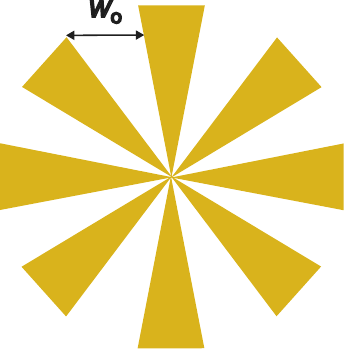}
\caption{Schematic representation of the simulated object and of the corresponding distance between features $w_o$}
\label{fig.obejctSize}
\end{figure}

An example of an object simulated in this study is schematically represented in figure \ref{fig.obejctSize}. It is a Siemens star test pattern with 50\% transmission and \qty{-1}{\radian} of phase shift. The spokes are generated by dividing a circle into equal angular intervals, whose amplitude controls the size of the object features. Only the upper left quadrant of the sample was kept during free-space propagation, to reduce the field of view ($250\times250$ \unit{\micro\meter\squared}) and therefore the required computational power. The spoke size $w_o$ was estimated at the edge of the star, along the horizontal/vertical scanning direction for vertical/horizontal spokes respectively, as shown in figure \ref{fig.obejctSize}. The value $w_o$ given in the text is therefore just an approximation and varies with the tilt of the spokes.

\section{Reconstruction: effect of step size}
\begin{figure}[htbp]
\centering
\includegraphics[scale = 0.3]{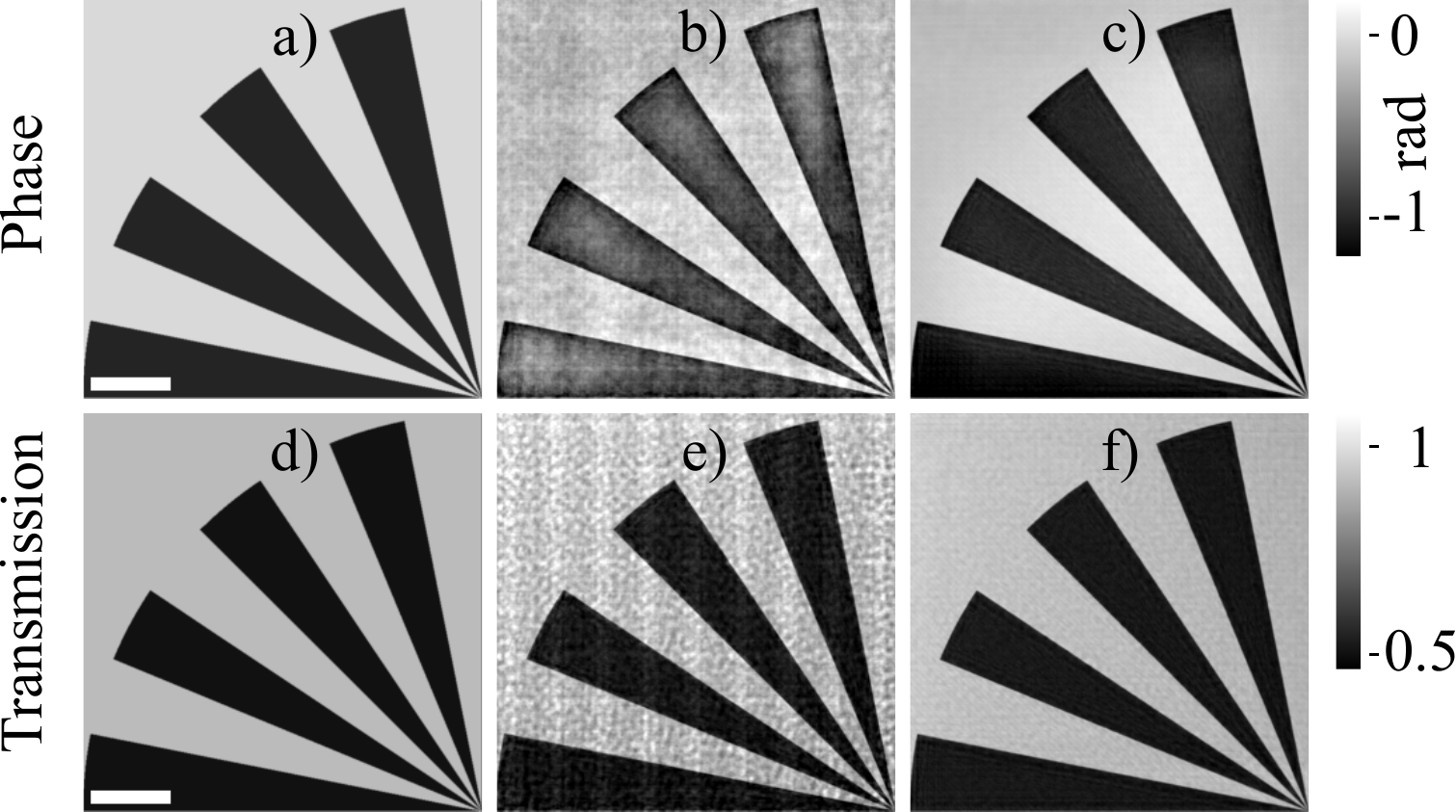}
\caption{Object phase (a-c) and transmission (d-f) reconstructed from simulations. (a,d) ground truth object ($O_1$). (b,e) reconstruction obtained from $d=$\qty{10}{\centi\meter} and $\Delta_{step}=$ \qty{28}{\micro\meter}. (c,f) same as (b,e) with $\Delta_{step}=$ \qty{4}{\micro\meter}. (b,e) correspond to a scanning step larger than $W$ and show artefacts in the reconstructed phase. The scale bars correspond to \qty{50}{\micro\meter}.}
\label{fig.figure_artefact_step}
\end{figure}
Figure \ref{fig.figure_artefact_step} represents the effect of the step size on the reconstructed sample's transmission and phase. Figures \ref{fig.figure_artefact_step} (a-c) represent the phase of the ground truth object and its reconstructions obtained with a step size of \qty{28}{\micro\meter} and \qty{4}{\micro\meter}, respectively. The propagation distance d=$\qty{10}{\centi\meter}$ and the pixel size \qty{0.42}{\micro\meter} provide a diffraction window $W$ of \qty{18.5}{\micro\meter}. In (b) the step size is greater than $W$ and the phase object appears hollow in between the edges of the spokes, where the phase tends towards the background value. Figures \ref{fig.figure_artefact_step} (d-f) show the transmission corresponding to (a-c). The transmission is always correctly recovered, even if background artefacts appear in (e), associated to a non-converging phase reconstruction in (b)

\section{Experimental results}
Hollow artefact due to violation of the NFP sampling condition 1 and 2 were observed also in experimental data. We report below the results of an experiment conducted using the Plasma X-ray Source (PXS) in the experimental area E1 of the ELI Beamlines facility of the Extreme Light Infrastructure ERIC, in Czech Republic. 

\begin{figure}[htbp]
\centering\includegraphics[scale=0.25]{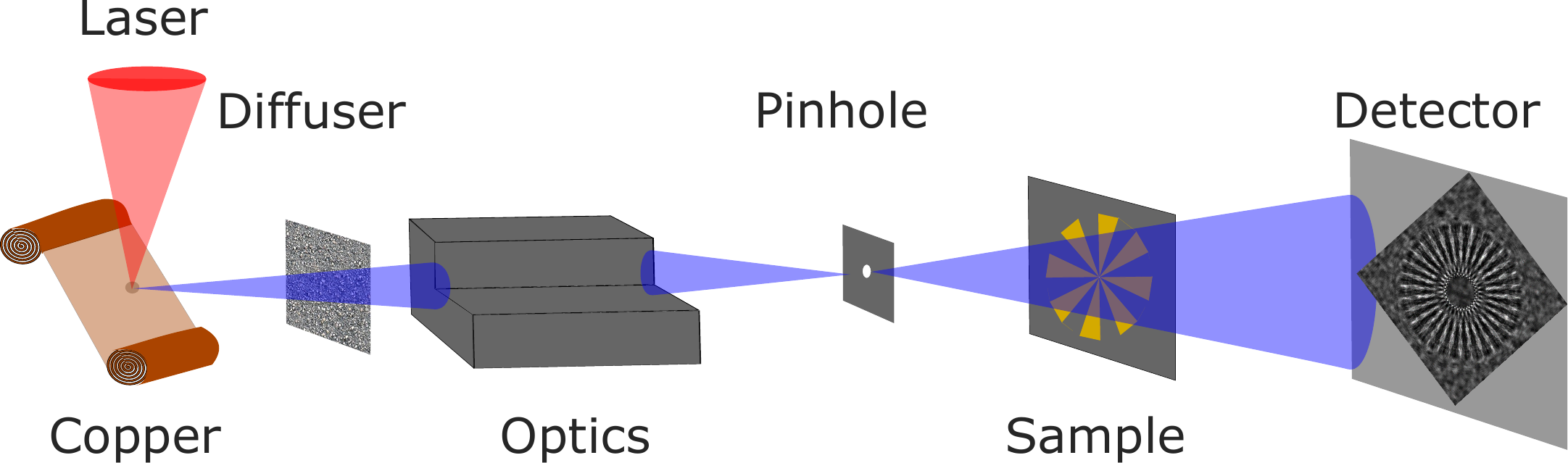}
\caption{Schematic of the experimental NFP setup at ELI Beamlines (not in scale).}
\label{fig.ELI_setup}
\end{figure}

The source is based on a \qty{20}{\milli\joule}, sub-20 \unit{\femto\second}, \qty{1}{\kilo\hertz} laser interacting with a copper tape to generate Cu $\mathrm{K}_{\alpha}$ emission at \qty{8}{\kilo\electronvolt} with sub-\unit{\pico\second} pulses. The experimental setup is shown in figure~\ref{fig.ELI_setup}. 
A foil of sandpaper, P200 grit, was used as a modulator to structure the beam and  placed between the X-ray source and a multi-layer Montel mirror monochromator optimised for the $\mathrm{K}_{\alpha}$ emission line at \qty{8.04}{\kilo\electronvolt}. The beam was focused down to an illumination size of \qty{\sim 100}{\micro\meter}, with a divergence of \qty{4.9}{\milli\radian}. A pinhole aperture with \qty{5}{\micro\meter} diameter was placed on the focus of the Montel mirror, to create a secondary coherent source by selecting the coherent fraction. The detector, an iKon L CCD camera (Oxford Instruments Andor Ltd, Belfast, UK) with \qty{13.5}{\micro\meter} pixel size, was placed at \qty{1}{\meter} from the pinhole: this distance was limited by the size of the experimental chamber. The test sample was a MicroCT Resolution Test Target (XRNanotech GmbH, Villigen, Switzerland), featuring a \qty{1.5}{\micro\meter} thick Au Siemens Star, with maximum spacing of \qty{50}{\micro\meter}. The sample was positioned at \qty{125}{\milli\meter} from the pinhole, corresponding to a magnification factor of 8. The illuminated area, at the sample position, was a square with a size of approximately \qty{600}{\micro\meter}. The effective pixel size for the given configuration was \qty{1.69}{\micro\meter}.

Following the previous NFP characterisation \cite{clare_characterization_2015}, scans were planned over 16 positions. In this first proof-of-principle test, a $4\times4$ regular grid was chosen for simplicity. During the measurements, the average life-time of the copper target limited  the number of acquired points to $3\times2$: replacing the copper target modified the illuminating probe, thus hindering a continuation of the scan. A scanning step of \qty{20}{\micro\meter} was used, of the order of the features in the test target, to ensure enough diversity between measurements while keeping a large overlap.
At each position, 150 images were acquired, with \qty{3}{\second} integration time. This resulted in approximately 10 minutes of acquisition per scanning position, taking into account also the readout time of the camera.  
The 150 images were averaged and background corrected prior to reconstruction. The reconstruction was performed using the ePIE algorithm implemented in PtyREX \cite{batey_th_2014}. An optimisation for the scanning positions around the nominal grid nodes was included \cite{maiden_annealing_2012}, as well as a correction for angular variation in the incoming beam, which can compensate for small variations in the probe, by adding an additional degree of freedom for each scan position \cite{batey_ffp_lmj_2021}.

\begin{figure}[h]
\centering\includegraphics[scale = 0.5]{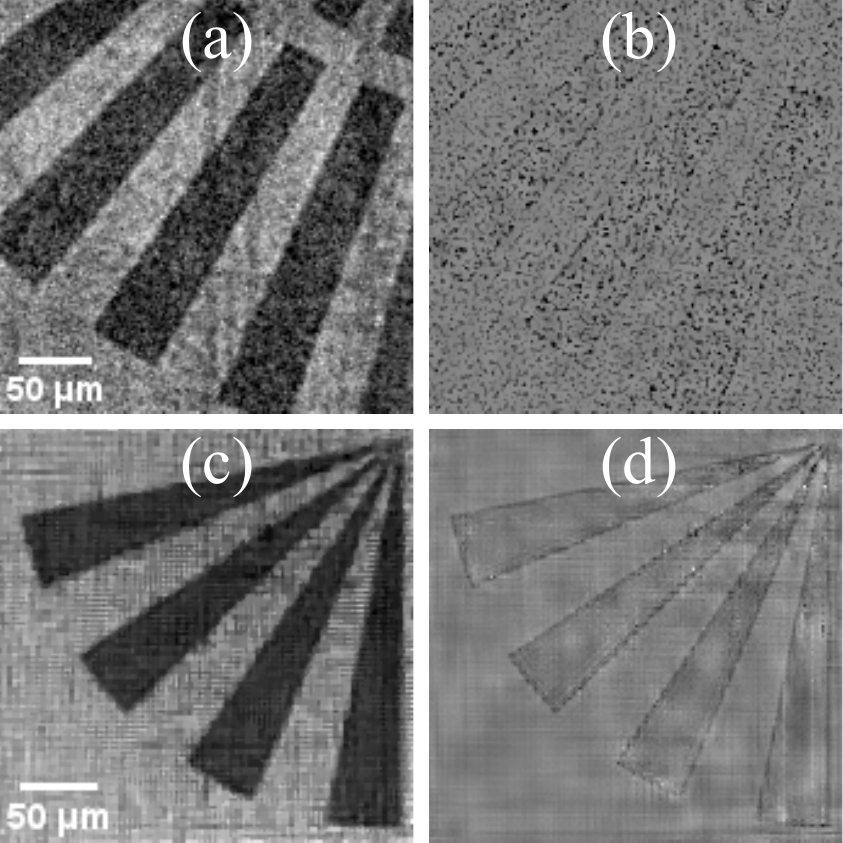}
\caption{Transmission (a) and phase (b) reconstructed from one of the ptychography scans acquired at ELI Beamlines. Transmission (c) and phase (d) reconstructed from a simulated scan, reproducing the same experimental conditions. Undersampling in the step size hinders the reconstruction of the phase.}
\label{fig.results_ELI}
\end{figure}

The retrieved sample's transmission and phase are represented in figure \ref{fig.results_ELI}-a and b  respectively. While a signal could be retrieved for the absorption, the phase is not recovered. The experimental configuration corresponded to a diffraction window $W$ of approximately \qty{5}{\micro\meter}, one fourth of the step size $\Delta_{step}=$ \qty{20}{\micro\meter}. The measurement was therefore characterised by severe undersampling in the step size. Figures \ref{fig.results_ELI}-c and d represent the transmission and phase of a Siemens Star, obtained by simulating a scan with the same experimental conditions. Only the number of points was changed and increased to $4 \times 4$. As for the experimental dataset, only the transmission is recovered, while the phase appears only at the edges of the spokes.

\bibliography{nfp}
\bibliographystyle{ieeetr}